\documentclass[twocolumn,prl,showpacs]{revtex4}

\usepackage{graphicx}
\usepackage{bm}

\begin{document}

\title{Temperature dependent
orbital degree of freedom in a bilayer manganite by magnetic Compton
scattering}

\author{ 
Yinwan Li$^{1,2}$, P. A. Montano$^{1,3}$, J.F. Mitchell$^2$,
B. Barbiellini$^4$, P. E. Mijnarends$^{4,5}$, 
S. Kaprzyk$^{4,6}$ and A. Bansil$^4$}

\affiliation{
$^1$Department of Physics, University of Illinois, Chicago IL 60680\\
$^2$Materials Science Division, Argonne National Laboratory, Argonne IL
60439\\
$^3$Materials Sciences and Engineering, U.S. Department of Energy, 1000
Independence Avenue, Washington DC 20585-1290\\
$^4$Physics Department, Northeastern University, Boston MA 02115 \\
$^5$Interfaculty Reactor Institute, Delft University of Technology,
2629 JB Delft, The Netherlands\\
$^6$Academy of Mining and Metallurgy AGH, 30059 Krak\'ow, Poland}

\date{\today}



\begin{abstract}

We have measured temperature-dependent magnetic Compton profiles (MCPs)
from a single crystal of La$_{1.2}$Sr$_{1.8}$Mn$_2$O$_7$. The MCPs, which
involved the scattering of circularly polarized x-rays, are in general
related to the momentum density of {\em all} the unpaired spins in the
system. Nevertheless, we show that when the x-ray scattering vector lies
along the [110] direction, the number of magnetic electrons of a {\em
specific} symmetry, i.e.  $d$-electrons of $x^2-y^2$ symmetry, yield a
distinct signature in the MCP, allowing us to monitor substantial changes
in the occupancy of the $d_{x^2-y^2}$ states over the investigated
temperature range of 5-200K. This study indicates that magnetic Compton
scattering can provide a powerful window on the properties of specific
magnetic electrons in complex materials.

\end{abstract}

\maketitle

Layered manganites ((La$_{1-x}$Sr$_x$MnO$_3$)$_{n}$SrO) have been the
subject of great current interest not only because they exhibit the   
colossal magnetoresistance (CMR) effect \cite{reviews,moritomo96}, but  
also because these systems display a wide variety of magnetic properties
\cite{kubota99} and undergo phase transitions associated with changes in
the orbital degrees of freedom. Here we consider the application of the 
x-ray scattering spectroscopy in the deeply inelastic or so-called   
Compton scattering regime (i.e. involving a large transfer of energy and
momentum in the scattering process) in order to gain a handle on the
properties of magnetic electrons in the manganites.

The scattering cross-section from magnetic electrons is
typically several orders of magnitude smaller than in the more
conventional
charge scattering channel, and therefore, magnetic Compton scattering
experiments have become feasible on complex materials only in the last few
years via the use of circularly polarized light at the
high energy, high intensity synchrotron sources \cite{mccarthy}.
Previous magnetic Compton scattering work 
on the manganites, which has been limited to the double layer
La-manganite, has focused on the case 
where the x-ray scattering vector lies along
the [100] or the [001] crystal direction 
\cite{li01,montano03,koizumi01,koizumi03}. 
The shape of the associated
[100] and [001] MCPs, however, is dominated by the $d$ electrons of $t_{2g}$
(i.e. $xy$, $yz$ and $zx$) symmetry. 
However, $t_{2g}$ levels lie $\sim$ 1 eV below the
Fermi energy and are thus relatively 'inert'--more central to the
understanding of the manganites is the behavior of the magnetic states of
$e_g$ (i.e. $x^2-y^2$ or $3z^2-r^2$) symmetry.

This article focuses on the case where the x-ray scattering vector is
chosen to lie along the [110] crystal direction. We show that the
resulting [110] MCP, when Fourier-transformed to real space, contains a
remarkable signature of the number of $d_{x^2-y^2}$ electrons through the
presence of a well-defined dip around 1 \AA. Along the [110] direction,
the shapes of various magnetic orbitals conspire in such a way as to
render the depth of the aforementioned dip quite insensitive to the
presence of other orbitals ($t_{2g}$ and $d_{3z^2-r^2}$). [110] MCPs in
La$_{1.2}$Sr$_{1.8}$Mn$_2$O$_7$ over the temperature range of 5K-200K are
presented. In order to help interpret these experimental results, we also
present the first all-electron computations of the MCP in the closely
related compound LaSr$_2$Mn$_2$O$_7$. From our analysis of the temperature
dependent MCP's, we adduce that the number of $d_{x^2-y^2}$ states
increases from about $0.3$ for each Mn atom at 5K to around $0.44$ at
160K, but then decreases rapidly to the low-$T$ value of $\sim0.3$ by
200K, suggesting that the system has undergone significant changes in the
electronic structure. We emphasize that the 7T magnetic field present
during the experiments makes the samples electronically homogeneous and
that this field is sufficiently strong to collapse polarons \cite{vasi99}.

The magnetic Compton profile, $J_{mag}(p_z)$, for momentum transfer along
the scattering vector $p_z$ is defined by
\begin{equation}
J_{mag}(p_z) =
            J_{\uparrow}(p_z) - J_{\downarrow}(p_z)~,
\label{eq1}             
\end{equation}
where $J_\uparrow$ ($J_\downarrow$) is the majority(minority) spin Compton
profile.
$J_{mag}$ can be expressed in terms of a double integral of the
spin density, $\Delta\rho({\mathbf p})$:
\begin{equation}
J_{mag}(p_z)= \int \int  \Delta\rho({\mathbf p}) dp_x dp_y,
\label{eq2}  
\end{equation}
where $\Delta\rho({\mathbf p})\equiv \rho_{\uparrow}({\mathbf p})
- \rho_{\downarrow}({\mathbf p})$.

MCP measurements were carried out using the elliptical multipole
wiggler at beamline 11-B at the Advanced Photon Source \cite{montano99}
on a high quality single crystal of
La$_{1.2}$Sr$_{1.8}$Mn$_2$O$_7$ using circularly polarized
photons. The sample is
$10 \times 5 \times 2 \mbox{ mm$^3$}$,
where the shortest side is along the crystal $c$-axis. The sample
was fixed to the holder with a specially made Al clip to avoid magnetic
contamination.   
The MCP's were obtained using magnetic fields of 3T and 7T along
the three high symmetry directions [001], [100], and [110] for four
different temperatures in each case: 5K, which is well below the Curie
temperature $T_c=$129K, and at 100K, 160K, and 200K, the last
temperature being well above $T_c$. The momentum resolution was $\sim 0.4$
a.u. $J_{\uparrow}(p_z)$ and
$J_{\downarrow}(p_z)$ were measured by flipping the photon polarization.
In this article, we focus on the [110] MCP and its temperature dependence
under a 7T field. We have also measured the MCP along [100] and [001]
and observed no significant variation in shape with temperature
\cite{li01,montano03}.

The electronic structure, magnetic momentum
density, and the MCPs along principal symmetry directions were computed
within an all-electron, fully charge and spin self-consistent
semi-relativistic KKR framework \cite{bansil99} for LaSr$_2$Mn$_2$O$_7$ in
the I4/mmm (No. 139) crystal structure \cite{seshadri97}. 
All computations were carried out to a high degree of accuracy, 
e.g. the
potentials were converged to better than $10^{-5}$ Ry. The momentum
density was computed over a fine mesh of $129.6 \times 10^6$ 
points within a sphere of radius 14.9 a.u. in ${\mathbf p}$ space
\cite{mijnarends95}.  This data set was used to evaluate various
spin-polarized projections and the MCP's.

Our computed electronic structure (for $x=0.5$)
is in good overall accord with two
relevant earlier studies \cite{deboer99,huang00}.  All calculations
agree on a nearly or wholly halfmetallic ferromagnetic band structure with
the Fermi level crossing Mn $d$ bands.  The $(\uparrow)$-spin bands
are metallic with a Fermi surface (FS) consisting of three sheets.
The ($\downarrow$)-spin FS sheet has the shape of a pillar.  
We obtain a spin-only magnetic moment
of $3.178$
$\mu_B$ on each Mn atom, in line with our experimental values of
3.4 $\mu_B$ at 7T and 3.1 $\mu_B$ at 3T, and of 3.0
$\mu_B$ for $x = 0.4$ \cite{mitchell97}. Most other atoms are found
to carry small moments of a few times
$10^{-2} \mu_B$. The total moment per formula unit is found to be 6.976
$\mu_B$, slightly lower than the values of 6.995 $\mu_B$ \cite{deboer99}
and 7.0 $\mu_B$ \cite{huang00} in earlier studies.

\begin{figure}
\begin{center}
\includegraphics[width=\hsize]{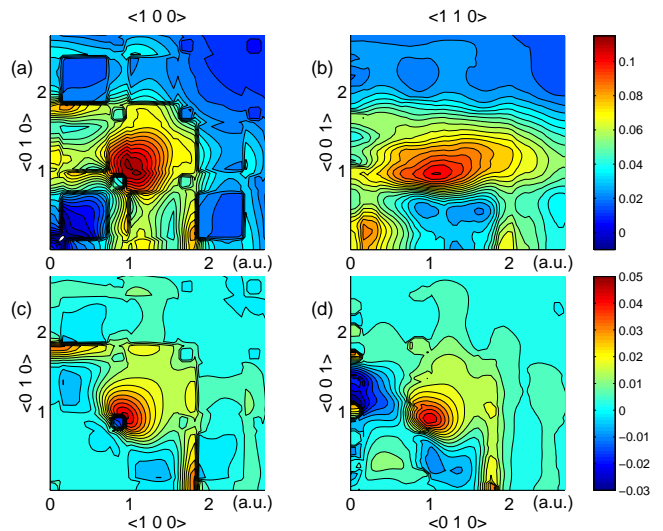}
\end{center}
\caption{
Theoretical 2D-projection (i.e. 1D-integral) of the magnetic momentum
density $\Delta \rho({\mathbf p})$ onto: (a) (001), and (b) (110) planes.  
(c) and (d)  give 2D-cuts through the 3D-density $\Delta \rho({\mathbf
p})$ 
in the $(p_z=0)$ and $(p_x=0)$ planes, respectively.}
\label{fig1}
\end{figure}

The interpretation of the momentum density and the MCP's is 
aided by Fig.~\ref{fig1}, which shows two typical 2D-projections (upper
frames) of the {\em magnetic} momentum density $\Delta \rho({\mathbf p})$, 
as well
as two orthogonal cuts through the origin of $\Delta \rho({\mathbf p})$
(lower frames). 
The (001) projection in Fig.~\ref{fig1}(a), for
instance, shows that the FS--seen by the presence of large and smaller
squarish features-- 
is essentially 2D in character. 
Underlying these FS features, the uncompensated
magnetic contribution yields a strong peak at $(p_x,p_y) \sim
(1.1,1.1)$ a.u. in the projection of (a). We have analyzed the origin of
this peak extensively by considering a series of cuts through
$\Delta\rho(\mathbf p)$ parallel to various high symmetry directions to
establish that this feature arises from wavefunctions of $t_{2g}$
symmetry.
Indeed, the (001) and the (100) cuts in (c) and (d), respectively, both
show a peak lying along the diagonal line as expected for an $xy$ or $yz$
type orbital.  

\begin{figure}
\begin{center}
\includegraphics[width=\hsize]{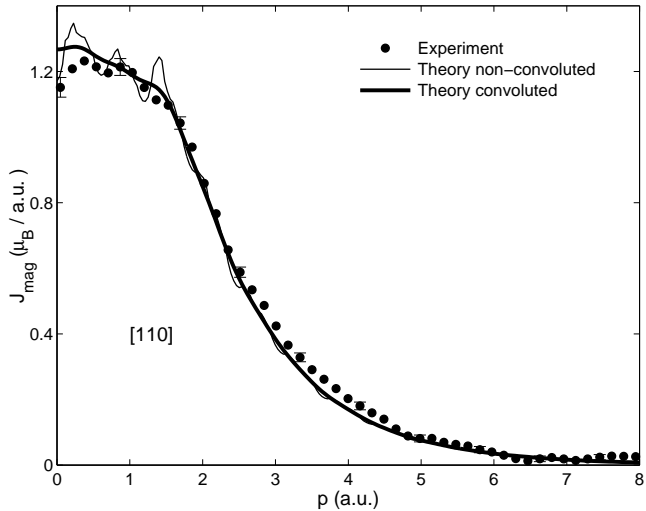}
\end{center}
\caption{Comparison of the theoretical and experimental [110] MCP, the
latter
taken at 5K under a 7T magnetic field along [110]. Theoretical curves are
shown with (thick line) and without (thin line) the effect of experimental
resolution of 0.4 a.u. (FWHM). All curves are normalized such that the
area
gives the measured magnetic moment of $3.4 \mu_B$/Mn atom. }
\label{fig2}
\end{figure}

Recalling from Eq. (\ref{eq2}) that the MCP involves a double integral or
equivalently a 1D-projection of $\Delta\rho({\mathbf p})$, the form of the
MCP's for momentum transfer along the principal symmetry directions can be
understood straightforwardly from the 2D-projections of Figs.~\ref{fig1}
(a) and (b).  For example, the [100] MCP represents a further projection
of the distribution of (a) on the horizontal [100]-axis, so that it
contains a $t_{2g}$ related peak around $1.1$ a.u., which is also the case
for the [001] MCP (as seen by projecting (b) on the vertical axis). Such a
peak is observed in our [100] and [001] MCP's and in those of
Ref.~\onlinecite{koizumi01}. These MCP's however are not shown here since
our focus in this article is on the [110] MCP. We see from (b) that the
region of high density is elongated along [110], so that the resulting
[110] MCP (Fig.~\ref{fig2}) is relatively flat unlike the [100] or the
[001] MCP's, particularly after the experimental resolution is included.  
Fig.~\ref{fig2} shows a remarkable level of agreement between theory and
experiment, some discrepancies notwithstanding \cite{agp}.
The structures
in the unbroadened theory MCP (thin solid line) are due partly to the FS,
but are washed out under the resolution of the present
experiment (0.4 a.u.). Higher resolution measurements would be most
interesting in gaining insight into
FS signatures in the momentum density.

\begin{figure}
\begin{center}
\includegraphics[width=\hsize]{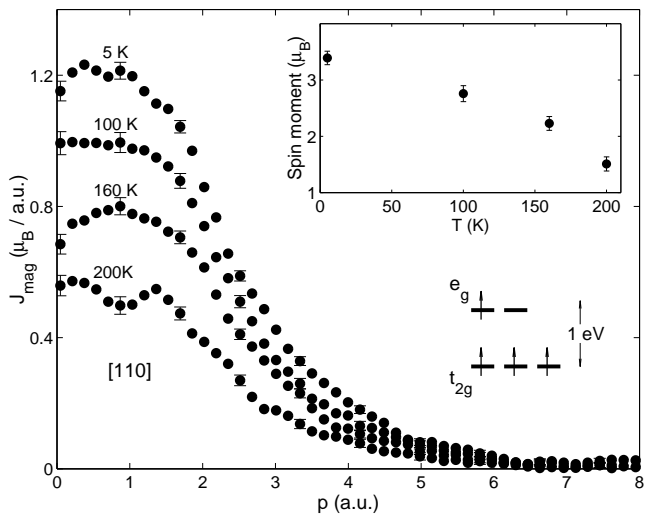}
\end{center}
\caption{ Temperature dependence of the [110] MCP under an external
magnetic field of 7T. Upper inset gives the corresponding spin magnetic
moments. A representative energy level diagram of the relevant magnetic
levels derived from our band structure computations is given in the lower
inset. }
\label{fig3}
\end{figure}

Fig.~\ref{fig3} presents the temperature dependence of the [110] MCP.  By
invoking a molecular orbital picture \cite{koizumi01}, substantial changes
observed in the shape of the [110] MCP with temperature can be understood.
Although such an analysis is not rigorous, it nevertheless provides a
handle on the important question of orbital occupancies. Assuming thus
that the $3.6$ Mn electrons essentially retain their atomic character, one
expects these to go into the up-spin states in accord with Hund's
first rule (see lower inset in Fig.~\ref{fig3}), with three electrons
occupying the up-spin $t_{2g}$ orbitals $(d_{xy}, d_{xz}, d_{yz})$ and
the remaining 0.6 electrons the $(d_{x^2-y^2}, d_{3z^2-r^2})$ orbitals.
The
total spin moment of course decreases with temperature following the
Curie-Weiss law as seen from the upper inset in Fig.~\ref{fig3}.
Note that the spin moment does not show anomalies 
observed at lower field 
\cite{kubota00}, confirming that the 
phase is electronically 
homogeneous under the field of 7T used in this work.

The energy level diagram in Fig.~\ref{fig3} makes it clear 
that the occupancy of
the $t_{2g}$ orbitals, which lie about $1$ eV below 
$E_F$--well above $kT$--
will change little with $T$, and 
that the variation with $T$ will occur mainly
through the redistribution of $0.6$ $e_g$ electrons 
between the $d_{x^2-y^2}$ and
$d_{3z^2-r^2}$ orbitals.
For this purpose, we consider the reciprocal form factor, 
$B({\mathbf r})$,  
which is defined as the Fourier transform of the spin momentum density
\cite{refbr}
\begin{equation}
B({\mathbf r})=\int \Delta \rho({\mathbf p})
\exp(-i{\mathbf p}\cdot{\mathbf r}) d{\mathbf p}~.
\label{eq3}
\end{equation}
We now express $\Delta\rho({\mathbf p})$ as a sum 
over the momentum density of the
molecular orbitals $\psi^{MO}_i({\mathbf p})$ 
of the magnetic electrons,
weighted by their occupancies $n_i$:
\begin{equation}
\Delta \rho({\mathbf p})=\sum_i n_i |\psi^{MO}_i({\mathbf p})|^2~.
\label{eq4}
\end{equation}
By transforming Eq. (\ref{eq4}) into real space, it is easily shown
that $B({\mathbf r})$ gives the autocorrelation of the magnetic 
orbitals, $\psi^{MO}_i({\mathbf r})$
\begin{equation}
B({\mathbf r})=\sum_{i}n_i\int \psi_i^{MO}({\mathbf s})
\psi_i^{*MO}({\mathbf s}+{\mathbf r}) d{\mathbf s}
\label{eq5}
\end{equation}
where the integrals are the standard two-center overlap functions.  
Alternately, $B({\mathbf r})$ along a given direction can be 
obtained directly by
taking the 1D-Fourier transform of the MCP along that direction by
comparing Eqs. (\ref{eq2}) and (\ref{eq3}).

\begin{figure}
\begin{center}
\includegraphics[width=\hsize]{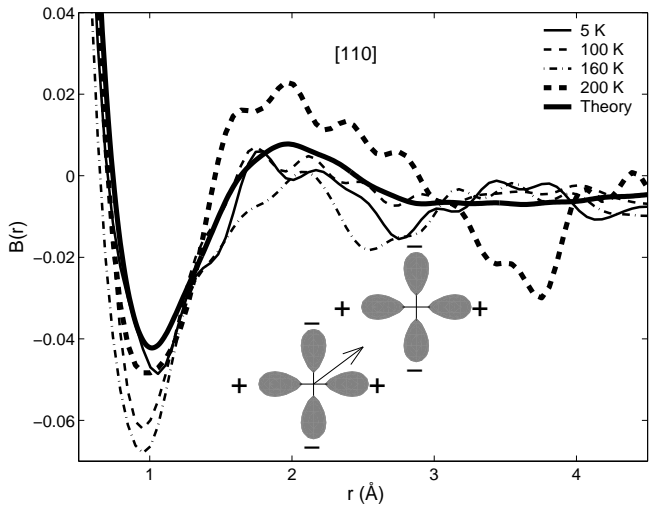}
\end{center}
\caption{
$B({r})$ results for $r$ along [110] obtained from the experimental and
the
broadened theoretical MCP's of Fig.~\ref{fig2}, and the MCP's at different
temperatures of Fig. 3. \cite{bmin} Inset: Schematic diagram of a
$d_{x^2-y^2}$
orbital
centered at the origin with lobes along [100] and [010] and the same
orbital displaced along [110]. }
\label{fig4}
\end{figure}

Fig.~\ref{fig4} presents the $B({\mathbf r})$ results along [110] obtained
from the MCP's of Figs.~\ref{fig2} and ~\ref{fig3}.  Notable features are
the presence of a dip at $r \approx 1$ \AA\  and oscillations at higher
distances.  As expected, the shape of the theory curve is in very good
accord with the 5K experimental data. Systematic changes in shape with
temperature can be interpreted by recalling that $B({\mathbf r})$ gives
the
autocorrelation of the magnetic orbitals. We consider the dip in
$B({\mathbf r})$ first. We have computed the overlap integral in
Eq. (5)  for the $t_{2g}$ and $e_g$ orbitals using  
Slater type orbitals (STO).  The key observation that emerges is that the
large dip in $B(r)$ along [110] is produced essentially by the
autocorrelation of the $d_{x^2-y^2}$ orbital.  This can be understood
qualitatively from the schematic picture of a $d_{x^2-y^2}$ orbital in
Fig.~\ref{fig4}.  
It is obvious that when this orbital is translated along
[110], the positive and negative lobes will overlap and yield a 
{\em negative} dip at a distance of the order of orbital dimensions. 
In contrast, the $d_{xy}$ and $d_{3z^2-r^2}$ 
orbitals give positive 
contributions near 1 \AA\ ; there also are small negative
contributions from $d_{xz}$ and $d_{yz}$ orbitals.
These results indicate that the size of the negative dip in $B({r})$ is
correlated with the $d_{x^2-y^2}$ occupancy.
Computations
in which all magnetic orbitals
are included show that
one can fit the minimum value,
$B_{min}$, of $B({r})$ along [110] by a
linear equation involving the occupancy $f$ of the $d_{x^2-y^2}$ orbital:
$B_{min}=af+b$, where $a=-0.130$ and $b=0.004$
\cite{bmin}.
Using this fit,
the $T$ dependent
data of Fig.~\ref{fig4} gives $f$ values of $0.38$, $0.48$, $0.53$ and
$0.38$, corresponding to 5, 100, 160 and 200K. Thus,
at low $T$, the $d_{3z^2-r^2}$ occupancy is $0.22$. By 160K, 
this number decreases to $0.07$, but then increases
relatively rapidly to the low-$T$ value by 200K.
These variations
correlate remarkably well with those in the apical oxygen distance,
$D_{apical}$, in the MnO$_6$ octahedra. Ref. \cite{mitchell97} reports
that $D_{apical}$ decreases up to $\approx$ 120K, but then begins to
increase quite rapidly.  A decrease in $D_{apical}$ is expected to raise
the energy of the $d_{3z^2-r^2}$ levels\ and to reduce their occupancy or,
equivalently, to increase the $d_{x^2-y^2}$ population, consistent with
earlier observations.

Turning to the behavior of $B(r)$ beyond the dip, Fig.~\ref{fig4}
shows that for $r \gtrsim 1.5$ \AA\  there is little overall change in
shape up to 160K. The 200K curve however displays striking differences in
that the broad feature around 2\AA\  develops a significantly larger
amplitude accompanied by the appearance of a new dip at $\approx 3.5$ \AA.
We emphasize that sizable values of $B({\mathbf r})$ beyond atomic
dimensions are a hallmark of electronic states extending over larger
distances as a result of the mixing of Mn and O orbitals in
the MnO$_2$ planes.  
The fact that at 200K $B(r)$ exhibits a
dramatic change in shape in that the dip reverts to the low-$T$ value and
strong oscillations appear at higher distances, indicates that the system
has undergone some orbital modification in which the oxygen
wavefunction character plays
an important role. We should keep in mind,
however, that, at the doping level of our sample, the system lies
near a multi-critical point in the phase diagram, so that
pin-pointing the detailed nature of the aforementioned orbital
change will generally be difficult. 

In conclusion, our study shows that magnetic Compton scattering spectra
taken under a field of 7T allow access to the properties of the
electronically
homogeneous phase of the manganite over a wide range of
temperatures. 
This strong magnetic field also collapses polarons.
The shape of the spectrum for momentum transfer along the
[110] direction contains a remarkable signature of the occupancy of the
$d_{x^2-y^2}$ electrons. These results indicate that magnetic Compton
scattering can provide a powerful new spectroscopic window for
investigating orbital physics and magnetic electrons in complex materials.

We acknowledge important discussions with Robert Markiewicz. This work is
supported by the US Department of Energy contract DE-AC03-76SF00098, and
benefited from the allocation of supercomputer time at NERSC, Northeastern
University's Advanced Scientific Computation Center (ASCC), and the
Stichting NCF (Foundation National Computer Facilities).



\begin{thebibliography}{99}

\bibitem{reviews}
  N. Mathur and P.
  Littlewood, Physics Today, January 2003, p. 25;
  Y. Tokura, Physics Today, July 2003, p. 50.

\bibitem{moritomo96}
  Y. Moritomo {\em et al.}, 
  Nature {\bf 380}, 141 (1996).

\bibitem{kubota99}
  M. Kubota {\em et al.},
  J. Phys. Chem. Solids {\bf 60}, 1161 (1999).
  

\bibitem{mccarthy} 
J.E. McCarthy {\em et al.}, J. Synchrotron Rad. {\bf 4}, 102 (1997).

\bibitem{li01}
  Y. Li {\em et al.}, Bull. Am. Phys. Soc. {\bf 46}, 167 (2001).

\bibitem{montano03}
  P.A. Montano {\em et al.}, Bull. Am. Phys. Soc. {\bf 48}, 193 (2003).
  
\bibitem{koizumi01}
  A. Koizumi {\em et al.},
  Phys. Rev. Lett. {\bf 86}, 5589 (2001).
  
\bibitem{koizumi03}
A. Koizumi {\em et al.},
Phys. Rev. B  {\bf 69}, 060401(R) (2004).
  
\bibitem{vasi99}
L. Vasiliu-Doloc et al., Phys. Rev. Lett. {\bf 83}, 4393 (1999). 

\bibitem{montano99}
  P.A. Montano {\em et al.}, Proc. SPIE {\bf 3773}, 262 (1999).


\bibitem{bansil99}
A. Bansil {\em et al.}, Phys. Rev. B {\bf 60}, 13 396 (1999).

\bibitem{seshadri97}
  R. Seshadri {\em et al.},
  Solid State Commun. {\bf 101}, 453 (1997).
  
\bibitem{mijnarends95}
  P.E. Mijnarends and A. Bansil, in {\em Positron Spectroscopy of
  Solids},
  eds. A. Dupasquier and A.P. Mills
  (IOS Press, Amsterdam, 1995), p. 25. 

\bibitem{deboer99}
  P.K. de Boer and R.A. de Groot,
  Phys. Rev. B {\bf 60}, 10 758 (1999).

\bibitem{huang00}
  X.Y. Huang {\em et al.},
  Phys. Rev. B {\bf 62}, 13 318 (2000).
  

\bibitem{mitchell97}
  J.F. Mitchell {\em et al.},
  Phys. Rev. B {\bf 55}, 63 (1997).

\bibitem{agp}
Our computations do not include 
correlation corrections, see e.g.
B. Barbiellini and A. Bansil, 
J. Phys. Chem. Solids {\bf 62}, 2181 (2001).

\bibitem{kubota00}
M. Kubota {\em et al.}, J. Phys. Soc. Jpn. {\bf 69}, 1986 (2000).

\bibitem{refbr}
B. Barbiellini and A. Shukla, Phys. Rev. B {\bf 66},
235101 (2002).

\bibitem{bmin}
The formula for $B_{min}$ was obtained by choosing STOs
and an occupancy of $3$ for the $t_{2g}$ states, of
$f$ for the $d_{x^2-y^2}$ and $(0.6-f)$ for the
$d_{3z^2-r^2}$ states.
The STOs yield
$B(r)=(1+t+c_2t^2+...+c_6t^6)\exp(-t)$, where
$t=Z_{eff} r/3$, $Z_{eff}$
is the effective charge of the atomic potential
and the coefficients $c_i$ depend
on $f$ (see, e.g., calculations of overlap integrals by
M.P. Barnett, Int. J. Quant. Chem. {\bf 95}, 791 (2003)).
In order to analyze changes in shape, all curves in Fig. 4 are normalized
such that, $B(0)=1$. 

\end{thebibliography}
\end{document}